\documentclass[prl,aps,twocolumn,preprintnumbers,amssymb,nobibnotes,nofootinbib,superscriptaddress,raggedbottom,10pt]{revtex4-1}
\usepackage{hyperref,graphicx,array,multirow,color,amsmath,mathrsfs,titlesec,shuffle,tikz}
\usetikzlibrary{calc,decorations,decorations.markings,decorations.pathmorphing,decorations.pathreplacing,shapes.geometric,patterns,shapes,decorations.pathmorphing,arrows}
\definecolor{plane1}{rgb}{0.9,0.5,0.65}
\definecolor{plane2}{rgb}{0.65,0.7,0.9}

\tikzset{bridgedAmp/.style={fill=ampgrey,circle,draw=black,line width=1.,minimum size=20pt,text=black,inner sep=0}}
\definecolor{ampgrey}{rgb}{0.9,0.9,0.9}
\tikzset{prop/.style={black,line width=1,line cap=round,rounded corners=0.05pt}}
\tikzset{path/.style={line width=1,line cap=round,rounded corners=3pt,decoration={},postaction={decorate},rounded corners=8.0pt}}
\tikzset{arrow/.style={path,decoration={markings,mark connection node=connode,mark=at position 0.5 with {\node[transform shape,scale=0.24,shape={dart},dart tip angle=40,fill=hteal] (connode) {};}},postaction={decorate}}}
\tikzset{int/.style={prop,decoration={},postaction={decorate},rounded corners=2.0pt}}
\tikzset{bdot/.style={fill=black,circle,minimum size=1.75pt,inner sep=0}}
\tikzset{feint/.style={black,line width=0.5,line cap=round,rounded corners=0.05pt}}
\tikzset{ddot/.style={fill=black,circle,minimum size=3.5pt,inner sep=0}}

\newcommand{\forwardLimitFigure}{\begin{tikzpicture}[scale=1,baseline=-3.05,rotate=0]\node[bdot]at($(0,0)+(80:0.6)$){};\node[bdot]at($(0,0)+(60:0.6)$){};\node[bdot]at($(0,0)+(40:0.6)$){};\node[bdot]at($(0,0)+(-80:0.6)$){};\node[bdot]at($(0,0)+(-40:0.60)$){};\node[bdot]at($(0,0)+(-60:0.6)$){};
\draw[int](0,0)--($(0,0)+(-100:0.75)$);\draw[int](0,0)--($(0,0)+(-20:0.75)$);\draw[int](0,0)--($(0,0)+(100:0.75)$);\draw[int](0,0)--($(0,0)+(20:0.75)$);\node[]at($(0,0)+(-100:0.95)+(0,0.05)$) {$\hat{n}$};\node[]at($(0,0)+(100:0.95)-(0,0.0025)$) {$\fwboxR{0pt}{\phantom{\hat{\ell}_0j\equivR}\!\!1}\!\!\!\!\!\!$};\draw[int,hblue](0,0) to [bend right=40] (-1.0,0.30);\draw[int,hblue](0,0) to [bend right=-40] (-1.0,-0.30);\node[]at($(-1,0)+(-0.125,0.0025)$) {$\!\!\fwboxL{0pt}{\b{\hat{\ell_1}}}$};\node[] at (0,0){$\hat{j}$};\node[bridgedAmp] at (0,0) {$\mathcal{A}_{n+2}^{L{-}1}$};\end{tikzpicture}}

\newcommand{\bridgeZeroFigure}{\begin{tikzpicture}[scale=1,baseline=-3.05,rotate=0]\node[bdot]at($(0,1)+(90:0.5)$){};\node[bdot]at($(0,1)+(60:0.5)$){};\node[bdot]at($(0,1)+(120:0.5)$){};\node[bdot]at($(0,-1)-(90:0.5)$){};\node[bdot]at($(0,-1)-(60:0.5)$){};\node[bdot]at($(0,-1)-(120:0.5)$){};
\draw[int](0,1) to (-0,0.30);\draw[int](0,-1) to (0,-0.30);\draw[int](0,-1)--($(0,-1)+(-150:0.75)$);\draw[int](0,-1)--($(0,-1)+(-30:0.75)$);\draw[int](0,1)--($(0,1)+(150:0.75)$);\draw[int](0,1)--($(0,1)+(30:0.75)$);\node[]at($(0,-1)+(-150:0.95)+(0,0.05)$) {$\hat{n}$};\node[]at($(0,1)+(150:0.95)-(0,0.0425)$) {$\fwboxR{0pt}{\phantom{\hat{\ell}_0j\equivR}\!\!1}\!\!\!\!\!\!$};\node[]at($(0,1)+(30:0.9)-(0,0.025)$) {$\!\!\fwboxL{0pt}{\phantom{j{-}1}\phantom{\hat{\ell}_0}}$};\node[]at($(0,-1)+(-30:0.9)+(0,0.025)$) {$\!\!\fwboxL{0pt}{\phantom{j}}$};\node[] at (0,0){$\hat{j}$};\node[bridgedAmp] at (0,1) {$T$};\node[bridgedAmp] at (0,-1) {$B$};\end{tikzpicture}}

\newcommand{\bridgeOneFigure}{\begin{tikzpicture}[scale=1,baseline=-3.05,rotate=0]\node[bdot]at($(0,1)+(90:0.5)$){};\node[bdot]at($(0,1)+(60:0.5)$){};\node[bdot]at($(0,1)+(120:0.5)$){};\node[bdot]at($(0,-1)-(90:0.5)$){};\node[bdot]at($(0,-1)-(60:0.5)$){};\node[bdot]at($(0,-1)-(120:0.5)$){};
\draw[int](0,1) to [bend left=-20] (-0.35,0.30);\draw[int](0,1) to [bend right=-20] (0.35,0.30);\draw[int](0,-1) to [bend right=20] (0.35,-0.30);\draw[int](0,-1) to [bend right=-20] (-0.35,-0.30);\draw[int](0,-1)--($(0,-1)+(-150:0.75)$);\draw[int](0,-1)--($(0,-1)+(-30:0.75)$);\draw[int](0,1)--($(0,1)+(150:0.75)$);\draw[int](0,1)--($(0,1)+(30:0.75)$);\node[]at($(0,-1)+(-150:0.95)+(0,0.05)$) {$\hat{n}$};\node[]at($(0,1)+(150:0.95)-(0,0.0425)$) {$\fwboxR{0pt}{\phantom{\hat{\ell}_0j\equivR}\!\!1}\!\!\!\!\!\!$};\node[]at($(0,1)+(30:0.9)-(0,0.025)$) {$\!\!\fwboxL{0pt}{\phantom{j{-}1}\phantom{\hat{\ell}_0}}$};\node[]at($(0,-1)+(-30:0.9)+(0,0.025)$) {$\!\!\fwboxL{0pt}{\phantom{j}}$}; \node[] at(-0.34,0) {$\hat{\ell}_1$};\node[] at (0.35,0){$\hat{j}$};\node[bridgedAmp] at (0,1) {$T$};\node[bridgedAmp] at (0,-1) {$B$};\end{tikzpicture}}

\newcommand{\bridgeLFigure}{\begin{tikzpicture}[scale=1,baseline=-3.05,rotate=0]\node[bdot]at($(0,1)+(90:0.5)$){};\node[bdot]at($(0,1)+(60:0.5)$){};\node[bdot]at($(0,1)+(120:0.5)$){};\node[bdot]at($(0,-1)-(90:0.5)$){};\node[bdot]at($(0,-1)-(60:0.5)$){};\node[bdot]at($(0,-1)-(120:0.5)$){};\draw[int](0,1) to [bend left=-40] (-1,0.30); \draw[int,hred](0,1) to [bend left=-30] (-0.45,0.30); \draw[int,hred](0,1) to [bend right=-30] (0.45,0.30);\draw[int](0,1) to [bend right=-40] (1,0.30);\draw[int](0,-1) to [bend right=40] (1.0,-0.30);\draw[int,hred](0,-1) to [bend right=30] (0.45,-0.30);\draw[int,hred](0,-1) to [bend right=-30] (-0.45,-0.30);\draw[int](0,-1) to [bend right=-40] (-1,-0.30);\draw[int](0,-1)--($(0,-1)+(-150:0.75)$);\draw[int](0,-1)--($(0,-1)+(-30:0.75)$);\draw[int](0,1)--($(0,1)+(150:0.75)$);\draw[int](0,1)--($(0,1)+(30:0.75)$);\node[]at($(0,-1)+(-150:0.95)+(0,0.05)$) {$\hat{n}$};\node[]at($(0,1)+(150:0.95)-(0,0.0425)$) {$\fwboxR{0pt}{\phantom{\hat{\ell}_0j\equivR}\!\!1}\!\!\!\!\!\!$};\node[]at($(0,1)+(30:0.9)-(0,0.025)$) {$\!\!\fwboxL{0pt}{\phantom{j{-}1}\phantom{\hat{\ell}_0}}$};\node[]at($(0,-1)+(-30:0.9)+(0,0.025)$) {$\!\!\fwboxL{0pt}{\phantom{j}}$}; \node[] at(-1,0) {$\hat{\ell}_1$};\node[] at(-0.45,0) {$\r{\hat{\ell}_2}$};\node[]at (0,0){$\r{\cdots}$};\node[]at(0.45,0) {$\,\,\,\r{\hat{\ell}_L}$};\node[] at (1,0){$\hat{j}$};\node[bridgedAmp] at (0,1) {$T$};\node[bridgedAmp] at (0,-1) {$B$};\end{tikzpicture}}

\newcommand{\bridgeLabelsFigure}{\begin{tikzpicture}[scale=1,baseline=-3.05,rotate=0]\node[bdot]at($(0,1)+(90:0.5)$){};\node[bdot]at($(0,1)+(60:0.5)$){};\node[bdot]at($(0,1)+(120:0.5)$){};\node[bdot]at($(0,-1)-(90:0.5)$){};\node[bdot]at($(0,-1)-(60:0.5)$){};\node[bdot]at($(0,-1)-(120:0.5)$){};\draw[int](0,1) to [bend left=-40] (-1,0.30); \draw[int](0,1) to [bend left=-30] (-0.45,0.30); \draw[int](0,1) to [bend right=-30] (0.45,0.30);\draw[int](0,1) to [bend right=-40] (1,0.30);\draw[int](0,-1) to [bend right=40] (1.0,-0.30);\draw[int](0,-1) to [bend right=30] (0.45,-0.30);\draw[int](0,-1) to [bend right=-30] (-0.45,-0.30);\draw[int](0,-1) to [bend right=-40] (-1,-0.30);\draw[int](0,-1)--($(0,-1)+(-160:0.75)$);\draw[int](0,-1)--($(0,-1)+(-20:0.75)$);\draw[int](0,1)--($(0,1)+(160:0.75)$);\draw[int](0,1)--($(0,1)+(20:0.75)$);\node[]at($(0,-1)+(-160:0.95)+(0,0.05)$) {$\hat{n}$};\node[]at($(0,1)+(160:0.95)-(0,0.0425)$) {$\fwboxR{0pt}{\phantom{j}\hat{\ell}_0\equivR\!\!1}\!\!\!\!\!\!$};\node[]at($(0,1)+(20:0.9)-(0,0.025)$) {$\!\!\fwboxL{0pt}{j{-}1\phantom{\hat{\ell}_0}}$};\node[]at($(0,-1)+(-20:0.9)+(0,0.025)$) {$\!\!\fwboxL{0pt}{j}$}; \node[] at(-1,0) {$\hat{\ell}_1$};\node[] at(-0.45,0) {${\hat{\ell}_2}$};\node[]at (0,0){${\cdots}$};\node[]at(0.45,0) {$\,\,\,{\hat{\ell}_q}$};\node[] at (1,0){$\hat{j}$};\node[bridgedAmp] at (0,1) {$T$};\node[bridgedAmp] at (0,-1) {$B$};\end{tikzpicture}}

\newcommand{\bridgeDefnFigure}{\begin{tikzpicture}[scale=1,baseline=-3.05,rotate=0]\node[bdot]at($(0,1)+(90:0.5)$){};\node[bdot]at($(0,1)+(60:0.5)$){};\node[bdot]at($(0,1)+(120:0.5)$){};\node[bdot]at($(0,-1)-(90:0.5)$){};\node[bdot]at($(0,-1)-(60:0.5)$){};\node[bdot]at($(0,-1)-(120:0.5)$){};\draw[int](0,1) to [bend left=-40] (-0.9,0.30); \draw[int,hred](0,1) to [bend left=-30] (-0.45,0.30); \draw[int,hred](0,1) to [bend right=-30] (0.45,0.30);\draw[int](0,1) to [bend right=-40] (0.9,0.30);\draw[int](0,-1) to [bend right=40] (0.9,-0.30);\draw[int,hred](0,-1) to [bend right=30] (0.45,-0.30);\draw[int,hred](0,-1) to [bend right=-30] (-0.45,-0.30);\draw[int](0,-1) to [bend right=-40] (-0.9,-0.30);\draw[int](0,-1)--($(0,-1)+(-160:0.75)$);\draw[int](0,-1)--($(0,-1)+(-20:0.75)$);\draw[int](0,1)--($(0,1)+(160:0.75)$);\draw[int](0,1)--($(0,1)+(20:0.75)$);\node[]at($(0,-1)+(-160:0.95)+(0,0.05)$) {$\hat{n}$};\node[]at($(0,1)+(160:0.95)-(0,0.0425)$) {$\fwboxR{0pt}{\phantom{j\hat{\ell}_0\equivR}\!\!1}\!\!\!\!\!\!$};\node[]at($(0,1)+(20:0.9)-(0,0.025)$) {$\!\!\fwboxL{0pt}{j{-}1\phantom{\hat{\ell}_0}}$};\node[]at($(0,-1)+(-20:0.9)+(0,0.025)$) {$\!\!\fwboxL{0pt}{j}$}; \node[] at(-0.9,0) {$\hat{\ell}_1$};\node[] at(-0.45,0) {$\r{\hat{\ell}_2}$};\node[]at (0,0){$\r{\cdots}$};\node[]at(0.45,0) {$\,\,\,\r{\hat{\ell}_q}$};\node[] at (0.9,0){$\hat{j}$};\node[bridgedAmp] at (0,1) {$T$};\node[bridgedAmp] at (0,-1) {$B$};\end{tikzpicture}}

\newcommand{\firstSequenceFigure}{\begin{tikzpicture}[scale=1,baseline=10.05,rotate=0]
\coordinate(p1)at(-0.5,1.2);
\coordinate(a1hat)at($(p1)+(240:1.5)$);
\coordinate(a10)at($(a1hat)+(170:1)$);
\coordinate(a1end)at($(a1hat)+(170:1.25)$);
\coordinate(b10)at($(a1hat)+(170+180:3.25)$);
\coordinate(b1end)at($(a1hat)+(170+180:3.45)$);
\coordinate(a2hat)at($(a1hat)+(10:2.5)$);
\coordinate(b20)at($(a2hat)+(85:1.4)$);
\coordinate(b2end)at($(a2hat)+(85:1.6)$);
\coordinate(a20)at($(a2hat)-(85:1.2)$);
\coordinate(a2end)at($(a2hat)-(85:1.4)$);
\coordinate(nhat)at($(a1hat)!0.45!(p1)$);
\coordinate(nend)at($(nhat)+(110:1.5)$);\coordinate(nm1end)at($(nhat)-(110:1.5)$);
\coordinate(n0)at($(nhat)+(110:1)$);\coordinate(nm1)at($(nhat)-(110:1.2)$);
\draw[int](nhat)--(nm1end);
\draw[int](a2end)--(a2hat);
\draw[draw=none,line width=0,fill=plane1,opacity=0.9](a1end)--(b1end)--($(p1)+(-10:2.8)$)--(p1);
\draw[int](b2end)--(a2hat);
\draw[int](a1end)--(b1end);\draw[int](nhat)--(nend);\draw[int,dashed](p1)--(a1hat);
\draw[feint,hteal]($(a1hat)-(10:1.25)$)--($(a1hat)+(10:3.05)$);
\draw[feint,black!50]($(a1hat)-(10-5*1:1.25)$)--($(a1hat)+(10-5*1:3.25)$);
\draw[feint,black!50]($(a1hat)-(10-5*2:1.25)$)--($(a1hat)+(10-5*2:3.25)$);
\draw[feint,black!50]($(a1hat)-(10-5*3:1.25)$)--($(a1hat)+(10-5*3:3.25)$);
\node[ddot,hteal!20!black,opacity=0.5]at($(a1hat)!.8!(a10)$){};
\node[ddot,hteal!40!black,opacity=0.5]at($(a1hat)!.6!(a10)$){};
\node[ddot,hteal!60!black,opacity=0.5]at($(a1hat)!.4!(a10)$){};
\node[ddot,hteal!80!black,opacity=0.5]at($(a1hat)!.2!(a10)$){};
\node[ddot]at(n0){};\node[ddot]at(nm1){};\node[ddot]at(p1){};\node[]at ($(p1)+(-0.15,0.15)$) {$1$};\node[]at ($(nhat)-(0.2,-0.05)$) {${\color{black}\hat{n}}$};\node[]at ($(n0)-(0.2,0.0)$) {$n$};\node[]at ($(nm1)+(0.4,0.05)$) {$n\,\mi1$};\node[]at ($(a1hat)+(0,-0.3)$) {${\color{black}\hat{\ell}_1}$};\node[]at ($(a2hat)+(0.24,0.05)$) {${\color{black}\hat{\ell}_2}$};\node[]at ($(a10)+(0,0.3)$) {${\color{black}A_1}$};\node[]at ($(a20)+(-0.25,0.0)$) {${\color{black}A_2}$};\node[]at ($(b20)+(0.25,0.0)$) {${\color{black}B_2}$};\node[]at ($(b10)+(0,-0.3)$) {${\color{black}B_1}$};
\node[ddot,hteal]at(a1hat){};\node[ddot,hteal]at(nhat){};\node[ddot,hteal]at(a2hat){};\node[ddot]at(b10){};\node[ddot]at(b20){};\node[ddot]at(a20){};\node[ddot]at(a10){};
\end{tikzpicture}}

\newcommand{\secondSequenceFigure}{\begin{tikzpicture}[scale=1,baseline=10.05,rotate=0]
\coordinate(p1)at(-0.5,1.2);
\coordinate(a1hat)at($(p1)+(240:1.5)$);
\coordinate(a10)at($(a1hat)+(170:1)$);
\coordinate(a1end)at($(a1hat)+(170:1.25)$);
\coordinate(b10)at($(a1hat)+(170+180:3.25)$);
\coordinate(b1end)at($(a1hat)+(170+180:3.45)$);
\coordinate(a2hat)at($(a1hat)+(10:2.5)$);
\coordinate(b20)at($(a2hat)+(85:1.4)$);
\coordinate(b2end)at($(a2hat)+(85:1.6)$);
\coordinate(a20)at($(a2hat)-(85:1.2)$);
\coordinate(a2end)at($(a2hat)-(85:1.4)$);
\coordinate(nhat)at($(a1hat)!0.45!(p1)$);
\coordinate(nend)at($(nhat)+(110:1.5)$);
\coordinate(nm1end)at($(nhat)-(110:1.5)$);
\coordinate(n0)at($(nhat)+(110:1)$);
\coordinate(nm1)at($(nhat)-(110:1.2)$);
\coordinate(a3hat)at($(a2hat)+(150:1.4)$);
\coordinate(a3end)at($(a3hat)+(180:1.75)$);  
\coordinate(b3end)at($(a3hat)+(180+180:1.95)$); 
\coordinate(a30)at($(a3hat)+(180:1.55)$);
\coordinate(b30)at($(a3hat)+(180+180:1.75)$); 
\draw[int](nhat)--(nm1end);
\draw[int](a2end)--(a2hat);
\draw[feint,black!75]($(a2hat)-(85+65*1/5:0.38)$)--(a2hat);
\draw[feint,black!75]($(a2hat)-(85+65*2/5:0.35)$)--(a2hat);
\draw[feint,black!75]($(a2hat)-(85+65*3/5:0.32)$)--(a2hat);
\draw[feint,black!75]($(a2hat)-(85+65*4/5:0.29)$)--(a2hat);
\draw[feint,hteal]($(a2hat)-(85+65*5/5:0.26)$)--(a2hat);
\draw[draw=none,line width=0,fill=plane1,opacity=0.9](a1end)--(b1end)--($(p1)+(-10:2.8)$)--(p1);
\node[]at ($(p1)+(-0.15,0.15)$) {$1$};\node[ddot]at(p1){};\draw[int](nhat)--(nend);\node[ddot,hteal]at(nhat){};
\draw[int](b3end)--(a3hat);
\node[]at ($(nhat)-(0.2,-0.05)$) {${\color{black}\hat{n}}$};
\draw[draw=none,line width=0,fill=plane2,opacity=0.9](a2hat)--(b20)--($(b20)+(187:2.675)$)--(a1hat);\draw[int](b2end)--(a2hat);
\draw[int](a1end)--(b1end);\draw[int](a3end)--(a3hat);
\draw[feint]($(a1hat)-(10:0.5)$)--($(a1hat)+(10:2.65)$);
\draw[feint,black!75](a2hat)--($(a2hat)+(85+65*1/5:1.6)$);
\draw[feint,black!75](a2hat)--($(a2hat)+(85+65*2/5:1.62)$);
\draw[feint,black!75](a2hat)--($(a2hat)+(85+65*3/5:1.64)$);
\draw[feint,black!75](a2hat)--($(a2hat)+(85+65*4/5:1.66)$);
\draw[feint,hteal](a2hat)--($(a2hat)+(85+65*5/5:1.68)$);
\node[]at ($(a1hat)+(0,-0.3)$) {${\color{black}\hat{\ell}_1}$};\node[]at ($(a3hat)+(0,-0.3)$) {${\color{black}\hat{\ell}_3}$};\node[]at ($(a2hat)+(0.3,0.05)$) {${\color{black}\hat{\ell}_2}$};
\node[]at ($(a30)+(0,0.3)$) {${\color{black}A_3}$};
\node[]at ($(b30)+(0,0.3)$) {${\color{black}B_3}$};
\node[ddot,hteal]at(a1hat){};\node[ddot]at(a30){};\node[ddot,hred]at(a3hat){};\node[ddot,hteal]at(a2hat){};
\node[ddot]at(b30){};
\end{tikzpicture}}

\newcommand{\fourLoopCutPathFigureA}{\begin{tikzpicture}[scale=1,baseline=-2.05,rotate=0]
\draw[int](0,1)--(0,-1);\draw[int](1,0)--(-1,0);\draw[int](1,1)--(1,-1);\draw[int](1,-1)--(-1,-1);\draw[int](-1,-1)--(-1,1);\draw[int](-1,1)--(1,1);
\draw[int](1,1)--($(1,1)+(45:0.4)$);\draw[int](1,-1)--($(1,-1)+(-45:0.4)$);\draw[int](-1,1)--($(-1,1)+(135:0.4)$);\draw[int](-1,-1)--($(-1,-1)+(-135:0.4)$);
\draw[int](0,1)--($(0,1)+(90:0.4)$);\draw[int](1,0)--($(1,0)+(0:0.4)$);\draw[int](0,-1)--($(0,-1)+(-90:0.4)$);
\node[bdot]at($(-0.75,1.2)$){};\node[bdot]at($(-0.5,1.2)$){};\node[bdot]at($(-0.25,1.2)$){};
\node[bdot]at($(-0.75,-1.2)$){};\node[bdot]at($(-0.5,-1.2)$){};\node[bdot]at($(-0.25,-1.2)$){};
\draw[path,hteal](-3/2,-1/2)--(1/2,-1/2)--(1/2,1/2)--(-1/2,1/2)--(-3/2,1/2);
\node[transform shape,scale=0.24,shape={dart},dart tip angle=40,fill=hteal]at(-1.3,-1/2){};
\node[transform shape,rotate=180,scale=0.24,shape={dart},dart tip angle=40,fill=hteal]at(-1.3,1/2){};
\node[ddot]at(-1,-1/2){};\node[ddot]at(0,-1/2){};\node[ddot]at(1/2,0){};\node[ddot]at(0,1/2){};\node[ddot]at(-1,1/2){};
\node[anchor=north east,inner sep=1pt]at(-1,-1/2){$\hat{\ell}_1$};\node[anchor=south west,inner sep=1pt]at(0,1/2){$\hat{\ell}_4$};
\node[anchor=north west,inner sep=1pt]at(1/2,0){$\hat{\ell}_3$};
\node[anchor=north east,inner sep=1pt]at(0,-1/2){$\hat{\ell}_2$};
\node[anchor=south east,inner sep=1.5pt]at(-1,1/2){$\hat{j}$};
\node[anchor=south east,inner sep=1.5pt]at(-1.3,1.15){$1$};
\node[anchor= east,inner sep=1.5pt]at(-1.3,-1.3){$\hat{n}$};
\end{tikzpicture}}

\newcommand{\fourLoopCutPathFigureB}{\begin{tikzpicture}[scale=1,baseline=-2.05,rotate=0]
\draw[int](0,1)--(0,-1);\draw[int](1,0)--(-1,0);\draw[int](1,1)--(1,-1);\draw[int](1,-1)--(-1,-1);\draw[int](-1,-1)--(-1,1);\draw[int](-1,1)--(1,1);
\draw[int](1,1)--($(1,1)+(45:0.4)$);\draw[int](1,-1)--($(1,-1)+(-45:0.4)$);\draw[int](-1,1)--($(-1,1)+(135:0.4)$);\draw[int](-1,-1)--($(-1,-1)+(-135:0.4)$);
\draw[int](0,1)--($(0,1)+(90:0.4)$);\draw[int](1,0)--($(1,0)+(0:0.4)$);\draw[int](0,-1)--($(0,-1)+(-90:0.4)$);
\node[bdot]at($(-0.75,1.2)$){};\node[bdot]at($(-0.5,1.2)$){};\node[bdot]at($(-0.25,1.2)$){};
\node[bdot]at($(-0.75,-1.2)$){};\node[bdot]at($(-0.5,-1.2)$){};\node[bdot]at($(-0.25,-1.2)$){};
\draw[path,hteal](-3/2,-1/2)--(1/2,-1/2)--(1/2,1/2)--(-1/2,1/2)--(-3/2,1/2);
\node[transform shape,rotate=180,scale=0.24,shape={dart},dart tip angle=40,fill=hteal]at(-1.3,-1/2){};
\node[transform shape,rotate=0,scale=0.24,shape={dart},dart tip angle=40,fill=hteal]at(-1.3,1/2){};
\node[ddot]at(-1,-1/2){};\node[ddot]at(0,-1/2){};\node[ddot]at(1/2,0){};\node[ddot]at(0,1/2){};\node[ddot]at(-1,1/2){};
\node[anchor=north east,inner sep=1pt]at(-1,-1/2){$\hat{j}$};\node[anchor=south west,inner sep=1pt]at(0,1/2){$\hat{\ell}_2$};
\node[anchor=north west,inner sep=1pt]at(1/2,0){$\hat{\ell}_3$};
\node[anchor=north east,inner sep=1pt]at(0,-1/2){$\hat{\ell}_4$};
\node[anchor=south east,inner sep=1.5pt]at(-1,1/2){$\hat{\ell}_1$};
\node[anchor=south east,inner sep=1.5pt]at(-1.3,1.15){$1$};
\node[anchor= east,inner sep=1.5pt]at(-1.3,-1.3){$\hat{n}$};
\end{tikzpicture}}

\newcommand{\fourLoopCutPathFigureC}{\begin{tikzpicture}[scale=1,baseline=-2.05,rotate=0]
\draw[int](0,1)--(0,-1);\draw[int](1,0)--(-1,0);\draw[int](1,1)--(1,-1);\draw[int](1,-1)--(-1,-1);\draw[int](-1,-1)--(-1,1);\draw[int](-1,1)--(1,1);
\draw[int](1,1)--($(1,1)+(45:0.4)$);\draw[int](1,-1)--($(1,-1)+(-45:0.4)$);\draw[int](-1,1)--($(-1,1)+(135:0.4)$);\draw[int](-1,-1)--($(-1,-1)+(-135:0.4)$);
\draw[int](0,1)--($(0,1)+(90:0.4)$);\draw[int](1,0)--($(1,0)+(0:0.4)$);\draw[int](0,-1)--($(0,-1)+(-90:0.4)$);
\node[bdot]at($(-0.75,1.2)$){};\node[bdot]at($(-0.5,1.2)$){};\node[bdot]at($(-0.25,1.2)$){};
\node[bdot]at($(-0.75,-1.2)$){};\node[bdot]at($(-0.5,-1.2)$){};\node[bdot]at($(-0.25,-1.2)$){};
\draw[path,hteal](-3/2,1/2)--(1/2,1/2)--(1/2,-1/2)--(-1/2,-1/2)--(-1/2,1/8);
\node[transform shape,scale=0.24,shape={dart},dart tip angle=40,fill=hteal]at(-1.3,1/2){};
\node[ddot]at(-1,1/2){};\node[ddot]at(0,1/2){};\node[ddot]at(1/2,0){};\node[ddot]at(0,-1/2){};\node[ddot]at(-1/2,0){};
\node[anchor=south east,inner sep=1pt]at(-1,1/2){$\hat{\ell}_1$};\node[anchor=south west,inner sep=1pt]at(0,1/2){$\hat{\ell}_2$};
\node[anchor=north west,inner sep=1pt]at(1/2,0){$\hat{\ell}_3$};
\node[anchor=north east,inner sep=1pt]at(0,-1/2){$\hat{\ell}_4$};
\node[anchor=south east,inner sep=1.5pt]at(-1/2,0){$\hat{j}$};
\node[anchor=south east,inner sep=1.5pt]at(-1.3,1.15){$1$};
\node[anchor= east,inner sep=1.5pt]at(-1.3,-1.3){$\hat{n}$};
\end{tikzpicture}}

\newcommand{\eightLoopCutPathFigure}{\begin{tikzpicture}[scale=0.8,baseline=-2.05,rotate=0]
\draw[int](0,1.92)--(0,-1.92);
\draw[int](1,0.45)--(-1,0.45);
\draw[int](1,-0.45)--(-1,-0.45);
\draw[int](1,-0.45)--(1,0.45);
\draw[int](-1,-0.45)--(-1,0.45);
\draw[int]($(-1,-0.45)+(-135:1.4)$)--($(1,-0.45)+(-45:1.4)$);
\draw[int]($(1,-0.45)+(-45:1.4)$)--($(1,0.45)+(45:1.4)$);
\draw[int]($(1,0.45)+(45:1.4)$)--($(-1,0.45)+(135:1.4)$);\draw[int]($(-1,0.45)+(135:1.4)$)--($(-1,-0.45)+(-135:1.4)$);
\draw[int](-1,-0.45)--($(-1,-0.45)+(-135:2)$);\draw[int](1,-0.45)--($(1,-0.45)+(-45:2)$);\draw[int](1,0.45)--($(1,0.45)+(45:2)$);\draw[int](-1,0.45)--($(-1,0.45)+(135:2)$);
\draw[path,hteal](-2.5,0)--(-1.5,0)--($(-1,-0.45)+(-135:0.7)$)--($(1,-0.45)+(-45:0.7)$)--(1.5,0)--(-.5,0)--(-.5,0.9)--(0.5,0.9)--(0.5,0.3);
\node[ddot]at($(-1.99,0)$){};
\node[ddot]at($(-1,-0.45)+(-135:0.585)$){};
\node[ddot]at($(0,-.945)$){};
\node[ddot]at($(1,-0.45)+(-45:0.585)$){};
\node[ddot]at($(1,0)$){};
\node[ddot]at($(0,0)$){};
\node[ddot]at($(-1/2,0.45)$){};
\node[ddot]at($(0,0.90)$){};
\node[ddot]at($(1/2,0.45)$){};
\node[anchor=north east, inner sep=1pt]at($(-1.99,0)$){$\hat{\ell}_1$};
\node[anchor=north, inner sep=1pt]at($(-1,-0.45)+(-135:0.585)-(0,0.05)$){$\hat{\ell}_2\hspace{-5pt}$};
\node[anchor=north east, inner sep=1pt]at($(0,-.945)$){$\hat{\ell}_3$};
\node[anchor=north, inner sep=1pt]at($(1,-0.45)+(-45:0.585)-(0,0.05)$){$\hspace{-0pt}\hat{\ell}_4$};
\node[anchor=south west, inner sep=1pt]at($(1,0)$){$\hat{\ell}_5$};
\node[anchor=north east, inner sep=0.5pt]at($(0,0)$){$\hat{\ell}_6$};
\node[anchor=south east, inner sep=1pt]at($(-1/2,0.45)$){$\hat{\ell}_7$};
\node[anchor=south east, inner sep=1pt]at($(0,0.90)$){$\hat{\ell}_8$};
\node[anchor=south west, inner sep=2pt]at($(1/2,0.45)$){$\hat{j}$};
\node[anchor=south east,inner sep=1.5pt]at(-2.4,1.75){$1$};
\node[anchor= east,inner sep=1.5pt]at(-2.4,-1.85){$\hat{n}$};
\end{tikzpicture}}

\titleformat{\section}{\centering\normalsize\normalfont\bf}{\thesection}{0em}{}
\hypersetup{pdftitle={},pdfcreator={},linkcolor=[rgb]{0.15,0.35,0.75},colorlinks=true,citecolor=[rgb]{0.675,0,0.2},urlcolor=[rgb]{0.15,0.35,0.65}}
\newcommand{\fwbox}[2]{\text{\makebox[#1][c]{$\hspace{-150pt}\displaystyle#2\hspace{-150pt}$}}}
\newcommand{\fwboxL}[2]{\text{\makebox[#1][l]{$#2$}}}
\newcommand{\fwboxR}[2]{\text{\makebox[#1][r]{$#2$}}}
\newcommand{\bigger}[1]{\raisebox{-0.95pt}{\scalebox{1.25}{$#1$}}}
\newcommand{\Bigger}[1]{\raisebox{-2.25pt}{\scalebox{1.75}{$#1$}}}
\renewcommand{\bar}{\overline}

\newcommand{\eq}[1]{\vspace{-3.5pt}\begin{equation}\hspace{2pt}#1\hspace{-0pt}\vspace{-3.5pt}\end{equation}}

\newcommand{\mi}{\raisebox{0.75pt}{\scalebox{0.75}{$\hspace{-1pt}\,-\,\hspace{-0.75pt}$}}}
\renewcommand{\pl}{\raisebox{0.75pt}{\scalebox{0.75}{$\hspace{-1pt}\,+\,\hspace{-0.75pt}$}}}
\newcommand{\ab}[1]{\langle #1\rangle}
\newcommand{\equivR}{\fwbox{13.5pt}{\hspace{-0pt}\fwboxR{0pt}{\raisebox{0.47pt}{\hspace{1.45pt}:\hspace{-3pt}}}=\fwboxL{0pt}{}}}
\newcommand{\equivL}{\fwbox{13.5pt}{\fwboxR{0pt}{}=\fwboxL{0pt}{\raisebox{0.47pt}{\hspace{-3pt}:\hspace{1.45pt}}}}}

\newcommand{\newcap}{\mathrm{\raisebox{0.75pt}{{$\,\bigcap\,$}}}}
\newcommand{\tcap}{\scalebox{1}{$\!\newcap\!$}}
\newcommand{\tncap}{\scalebox{0.8}{$\!\newcap\!$}}
\renewcommand{\hat}{\widehat}
\renewcommand{\phi}{\varphi}

\definecolor{hblue}{rgb}{0,0,0.575}
\definecolor{hred}{rgb}{0.575,0.0,0.225}
\definecolor{hgreen}{rgb}{0.075,0.4,0.125}
\definecolor{hteal}{rgb}{0.0,0.545,0.7451}
\definecolor{perm}{rgb}{0.1,0.45,0.85}

\renewcommand{\r}[1]{{\color{hred}#1}}
\renewcommand{\b}[1]{{\color{hblue}#1}}
\newcommand{\g}[1]{{\color{hgreen}#1}}

\begin{document}
\title{\texorpdfstring{Loops from Cuts\\[-22pt]~}{Loops from Cuts}}
\author{Jacob~L.~Bourjaily}
\affiliation{Institute for Gravitation and the Cosmos, Department of Physics,\\Pennsylvania State University, University Park, PA 16802, USA}
\author{Simon Caron-Huot}
\affiliation{Department of Physics, McGill University, 3600 Rue University, Montr\'eal, H3A 2T8, QC Canada}


\begin{abstract}
We derive novel recursion relations for all loop amplitude integrands of planar, maximally supersymmetric Yang-Mills theory in terms of unitarity-like `cuts' obtained via sequences of BCFW deformations in momentum-twistor space. \vspace{-10pt}
\end{abstract}
\maketitle

\vspace{-15pt}\section{Introduction}\label{introduction_section}\vspace{-14pt}
%
Much of the recent progress in our understanding of scattering amplitudes in perturbative quantum field theories has stemmed from powerful new techniques to efficiently compute amplitudes and loop-amplitude integrands at large multiplicity and/or loop-order. These techniques have provided a rich source of theoretical `data' from which a number of deep and important insights have been gleamed. For example, building on analytic insights into tree amplitudes (see e.g.~\cite{Elvang:2013cua}), the all-loop recursion relations for loop integrands discovered in \cite{ArkaniHamed:2010kv} led directly to a wealth of theoretical data, against which new ideas were rapidly developed and tested. 

The recursion relations described in \cite{ArkaniHamed:2010kv} generated loop amplitude integrands from the `forward-limits' of lower-loop amplitude integrands; although such forward-limits are believed to be well-defined for amplitudes in any supersymmetric field theory \cite{CaronHuot:2010zt}, they are known to involve considerable subtlety for less-than-maximal supersymmetry (especially for amplitudes represented via recursion) \mbox{\cite{Benincasa:2015zna,Benincasa:2016awv}}; and even for maximally supersymmetric Yang-Mills theory in the planar limit (`sYM'), forward limits prove an enormous source of computational inefficiency, greatly limiting their effective implementation. To date, they have been exploited for amplitudes only through two-loop order \mbox{\cite{Bourjaily:2010wh,Bourjaily:2013mma,Bourjaily:2015jna}}. 

In this note, we overcome this barrier by showing how sequences of BCFW recursions can be used to represent any $L$-loop, $n$-particle amplitude integrand of planar sYM, $\mathcal{A}_n^L$, in a form reminiscent of unitarity cuts:
\vspace{-5pt}\eq{\hspace{-60pt}\mathcal{A}_n^{L}\bigger{=}\fwboxL{5pt}{\bigger{\sum}}\hspace{-0pt}\bridgeZeroFigure\hspace{-5pt}\bigger{+}\hspace{-5pt}\bridgeOneFigure\hspace{-10pt}\bigger{+}\ldots\bigger{+}\bridgeLFigure\,.\hspace{-40pt}\label{schematic_cut_formula}}
Here, the first term corresponds to the ordinary `BCFW bridge', and the second the one-loop `kermit' terms described in \cite{Bourjaily:2013mma}. At two loops, a closed formula similar to (\ref{schematic_cut_formula})
was given in the appendix of \cite{Bourjaily:2015jna}, but it involved two sums over triple-products of amplitudes. Despite the similarity between \eqref{schematic_cut_formula} and unitarity cuts, the representation we derive features surprisingly novel types of cuts at three-loops and higher. 

Attached to this note, we have prepared an implementation of the representation (\ref{schematic_cut_formula}) in \textsc{Mathematica} which has been checked both internally and against a number of non-trivial examples and has used to compute a number of previously inaccessible loop amplitudes including: the $10$-particle N${}^3$MHV amplitude at 3 loops, the 8-particle N${}^2$MHV amplitude at 4 loops, and the $5$-particle MHV amplitude at 5 loops.

\vspace{-20pt}\section{BCFW recursion in Momentum-Twistor Space}\vspace{-14pt}
%

\begin{figure*}[t]
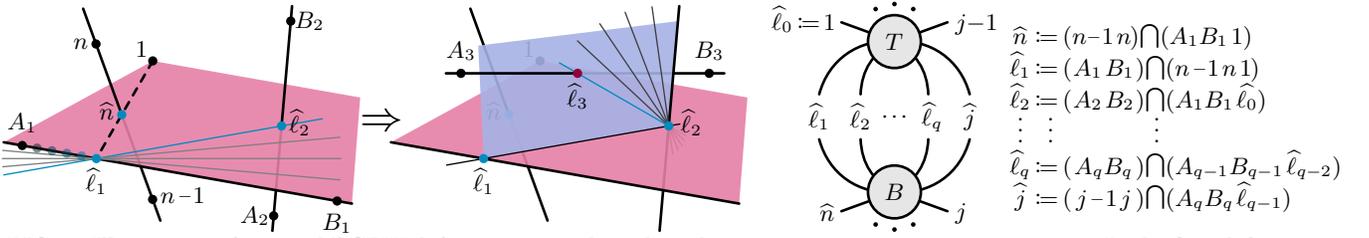
\fwboxR{300pt}{\hspace{-30pt}\firstSequenceFigure\hspace{0pt}$\Bigger{\Rightarrow}$\hspace{-5pt}\secondSequenceFigure
\hspace{25pt}$\fwboxL{100pt}{\hspace{-5pt}
\bridgeLabelsFigure\hspace{10pt}\begin{array}{@{}c@{}l@{$\!$}l@{}}\hat{n}&\equivR&(n\mi1\,n\hspace{-0.5pt})\tcap(A_1B_1\,1)\\
\hat{\ell}_1&\equivR&(\fwbox{24pt}{A_1\,B_1})\tcap(n\,\mi1\,n\,1\hspace{-0.5pt})\\
{\hat{\ell}_2}&\equivR&(\fwbox{24pt}{A_2\,B_2})\tcap(A_1B_1\,\hat{\ell_0})\\[-5pt]
{\vdots}&\;\;\,\,\vdots&\hspace{34.5pt}{\vdots}\\[-0pt]
{\hat{\ell}_q}&\equivR&(\fwbox{24pt}{A_qB_q})\tcap(A_{q{-}1}B_{q{-}1}\,\hat{\ell}_{q{-}2})\\
{\hat{j}}&\equivR&(\fwbox{24pt}{j\,\mi1\,j})\tcap(A_{q}B_{q}\,\hat{\ell}_{q{-}1})\\
\end{array}\hspace{-20pt}}$}\vspace{-12pt}\caption{Illustration of iterated BCFW deformations and resulting locations in momentum twistor space. In the first deformation, $z_n$ is translated by $z_{n{-}1}$ until it intersects the plane $(\ell_1\,1)$ at the point $\hat{n}\equivR\!(n\,\mi1\,n)\tncap(A_1\,B_1\,1)$ (cutting the local pole $\ab{\ell_1\,n\,1}$); in this configuration, we may translate $A_1$ to $\hat{\ell}_1\equivR\!(A_1\,B_1)\tncap(n\,\mi1\,n\,1)$. After this, we perform a second BCFW deformation by translating $B_1$ in the direction of $\hat{n}$ (so as to remain in the plane $(A_1\,B_1\,1)$) until it intersects the line $(A_2\,B_2)$ representing $\ell_2$ at the point $\hat{\ell}_2\equivR\!(A_2\,B_2)\tncap(A_1\,B_1\,1)$---which cuts the internal propagator $\ab{\ell_1\,\ell_2}$. In the second figure, this sequence is continued, by deforming $\ell_2$ in the plane $(A_2\,B_2\,\hat{\ell}_1)$ until it intersects the line $(A_3\,B_3)$ at the point $\hat{\ell}_3\equivR\!(A_3\,B_3)\tncap(A_2\,B_2\,\hat{\ell}_1)$, and so on.\\[-6pt] \label{cut_sequence_figure}}\end{figure*}

Our starting point will be the all-loop recursion of \cite{ArkaniHamed:2010kv}. Similar to the tree-level BCFW recursion \cite{BCF,BCFW}, it may be derived by shifting a pair of external momenta by a complex parameter while maintaining on-shell and momentum conservation constraints. In sYM (and for suitable helicity choices in non-supersymmetric theories), the amplitude vanishes at infinite momentum shift \cite{ArkaniHamed:2008yf} and therefore is fully determined by its poles at finite shifts. At tree-level, these poles represent factorization channels, the residues of which can be determined recursively as products of lower-point, on-shell amplitudes. For loop amplitude integrands, there are also poles corresponding to locations where internal propagators go on-shell, which can be described (if not easily computed) as a lower-loop amplitude with two additional momenta taken in the `forward-limit'.

The mechanics of on-shell recursion are simplified in momentum-twistor variables \cite{Hodges:2009hk} which simultaneously trivialize both momentum conservation and the on-shell condition for massless particles. Thus, an arbitrary set of momentum twistors $z_a\!\in\!\mathbb{C}^4/\text{GL}(1)$ will encode a set of momentum-conserving, massless four-momenta. This is achieved using the twistor map \cite{Penrose:1967wn} which associates the (projective) line $(a\mi1\,a)\equivR\!\mathrm{span}\{z_{a{-}1},z_a\}$ (cyclic labeling understood) in twistor space with a point $x_a$ in spacetime. Two points in spacetime will be light-like separated if their corresponding lines in twistor space intersect; as such, the points $x_a$ and $x_{a+1}$ are manifestly light-like separated because the corresponding lines in twistor space  $(a\,\mi1\,a)$ and $(a\,a\pl1)$ intersect at $z_a$. The collection of pairwise, null-separated points $\{x_a\}$ can be used to define a set of null momenta $p_a\equivR\!(x_{a{+}1}-x_a)$, for which momentum conservation is manifest.
For supersymmetric Yang-Mills theory each twistor $z_a$ can be upgraded to a projective vector in $\mathbb{C}^{4|\mathcal{N}}$ which captures the helicity dependence of scattering amplitudes.
These variables make sYM formulas particularly compact as they manifest its dual conformal symmetry \cite{Drummond:2006rz,Drummond:2009fd,Mason:2009qx}.

When two dual-momentum points $x_{\r{a}}$ and $x_{\b{b}}$ become like-like separated, their corresponding lines in momentum-twistor space $(\r{a\,\mi1\,a})$ and $(\b{b\,\mi1\,b})$ intersect; this corresponds to a situation where the four twistors $\{z_{\r{a-1}},z_{\r{a}},z_{\b{b-1}},z_{\b{b}}\}$ span a space of rank less than four. Defining the `four-bracket' $\ab{a\,b\,c\,d}\equivR\!\det\{z_a,z_b,z_c,z_d\}$, two lines intersect when $\ab{\r{a\,\mi\,1\,a}\,\b{b\,\mi1\,b}}{=}0$. 

For planar theories described in dual-momentum coordinates, physical poles must involve the sums of consecutive momenta $(x_{\r{a}}{-}x_{\b{b}})^2{=}(p_{\r{a}}{+}p_{a+1}{+}\ldots{+}p_{\b{b}-1})^2$, or involve some internal propagator $(x_{\b{\ell_i}}{-}x_{\r{b}})^2$ where $x_{\b{\ell_i}}$ represents an internal loop momentum---to be associated with the line $(\b{\ell_i})\equivR\!(\b{A_i\,B_i})$ in momentum twistors spanned by $z_{\b{A_i}},z_{\b{B_i}}$. 

Consider the BCFW deformation of (super)-twistors
\eq{z_n\!\mapsto \g{\bar{z}_n}(\r{\alpha})\equivR\!\g{z_n}{+}\r{\alpha}\,\g{z_{n{-}1}}\,;\label{general_shift}}
this deformation trivially preserves the on-shell condition and momentum conservation. Since the only \emph{line} deformed is $(\g{z_n,z_1})$---corresponding the dual point $x_1$---this is equivalent to BCFW-deforming the pair of momenta $p_n$ and $p_1$. Although for sYM there is no pole at infinite external momenta
there is a pole at $\r{\alpha}\!\to\!\infty$ corresponding to the factorization involving a three-particle
$\overline{\text{MHV}}$ amplitude (an `inverse soft-factor' in the language of \mbox{\cite{ArkaniHamed:2009dg,ArkaniHamed:2012nw}})---which, in momentum twistor space, is simply the lower-point amplitude involving twistors $\{z_1,\ldots,z_{n-1}\}$. 

The poles at finite values of $\r{\alpha}$ must involve the vanishing of either $\ab{\g{\bar{n}\,1}\,\b{j\,\mi1\,j}}$, a factorization, or of $\ab{\g{\bar{n}\,1}\,\b{A_i B_i}}$, a `forward-limit'. The locations of these poles are given by
\eq{\fwbox{0pt}{\hspace{-5pt}\g{\hat{n}}=\!(\g{n\,\mi1\,n})\tncap(\g{1}\,\b{j\,\mi1\,j})\quad\text{or}\quad\g{\hat{n}}=\!(\g{n\,\mi1\,n})\tncap(\g{1}\,\b{A_i\,B_i}),}\label{int_and_external_poles}}
where $(\r{a\,b})\tncap(\b{c\,d\,e})\equivR\!z_{\r{a}}\,\ab{\r{b}\,\b{c\,d\,e}}{+}z_{\r{b}}\,\ab{\b{c\,d\,e}\,\r{a}}$---or, equivalently, in the parameterization (\ref{general_shift}), at the locations 
\eq{\r{\alpha}=\frac{\ab{\g{n}\,\g{1}\,\b{j\mi1}\,\b{j}}}{\ab{\g{1}\,\b{j\mi1}\,\b{j}\,\g{n{-}1}}}\quad\text{or}\quad\r{\alpha}=\frac{\ab{\g{n}\,1\,\b{A_i}\,\b{B_i}}}{\ab{\g{1}\,\b{A_i}\,\b{B_i}\,\g{n{-}1}}}\,.}
Residues involving `external' lines $(\b{j\mi1}\,\b{j})$ correspond to factorization channels involving two lower-multiplicity amplitude integrands involving any distribution of the total $L$ loop momenta between them, represented by the first term of (\ref{schematic_cut_formula}). Concretely, these terms are given by 
\eq{\mathcal{A}_{n_T}^{L_T}\!\big(\g{1},\ldots,\b{j\mi1},\b{\hat{j}}\big)\mathcal{B}_0[\g{1}\,(\b{j\mi1\,j}),(\g{n\mi1\,n})]\mathcal{A}_{n_B}^{L_B}\!\big(\b{\hat{j}},\b{j},\ldots,\g{n\mi1},\g{\hat{n}}\big)\,\label{zero_bridge}}
where $L{=}L_B{+}L_T$, $\b{\hat{j}}\equivR(\b{j\mi1\,j})\tncap(\g{n\mi1\,n\,1})$.
Taking (as in \cite{ArkaniHamed:2010kv}) the amplitudes $\mathcal{A}$ to be the conventional amplitude (divided by the Parke-Taylor MHV super-amplitude),
the `bridge' $\mathcal{B}_0[\g{1},(\b{j\mi1,j}),(\g{n\mi1,n})]\equivR\mathcal{R}[\g{1},\b{j\mi1},\b{j},\g{n\mi1},\g{n}]$ is given by the `$R$-invariant' \cite{Drummond:2008vq} in momentum-twistor variables \cite{Hodges:2009hk}
\eq{\begin{split}&\mathcal{R}[a\,b\,c\,d\,e]\equivR\\
&\frac{\delta^{0|4}\!\big(\eta_{a}\ab{b\,c\,d\,e}{+}\eta_b\ab{c\,d\,e\,a}{+}\eta_c\ab{d\,e\,a\,b}{+}\eta_d\ab{e\,a\,b\,c}{+}\eta_e\ab{a\,b\,c\,d}\big)}{\ab{a\,b\,c\,d}\ab{b\,c\,d\,e}\ab{c\,d\,e\,a}\ab{d\,e\,a\,b}\ab{e\,a\,b\,c}}.\\[-16pt]\end{split}\nonumber}

The second type of pole---that involving an internal line $(\b{A_i}\,\b{B_i})$---corresponds to a one-lower-loop amplitude involving two additional particles taken in the forward limit as described in  \cite{ArkaniHamed:2010kv}; graphically, it corresponds to
\vspace{-10pt}\eq{\fwbox{0pt}{\forwardLimitFigure\equivR\underset{\b{A_1}\to\b{B_1}}{\text{FL}}\Big(\g{1},\ldots,\g{n\mi1},\g{\hat{n}},\b{A_1},\b{B_1}\Big)\,.}\vspace{-5pt}}
However, the computational complexities of analytically taking lower-loop amplitudes in the forward-limit have continued to stymie implementation or wide applications.

The loop integrand obtained in this fashion, after symmetrizing in the loop variables $x_{\b{\ell_i}}$, is a well-defined and gauge-invariant function
(as opposed to an equivalence class modulo shifts, which it normally is in a non-planar theory). It is conjectured to be the canonical volume form on the amplituhedron \cite{Arkani-Hamed:2013jha,Arkani-Hamed:2013kca}, and can be alternatively computed using a Wilson loop in twistor space \cite{Mason:2010yk}. By exploiting the duality between Wilson loops and scattering amplitudes in the sYM theory (see \cite{Alday:2007hr,Berkovits:2008ic,Beisert:2008iq}), it is also equal to a correlator of a null polygonal Wilson loops with Lagrangian insertions \cite{Caron-Huot:2010ryg} and can be obtained from the null limit of stress-tensor multiplet correlators \cite{Eden:2011yp,Eden:2011ku,Adamo:2011dq}. The latter has been discussed recently for massive amplitudes along the Coulomb branch \cite{Caron-Huot:2021usw}, but we focus here on the massless case where we benefit from the massless momentum twistor formalism.

\vspace{-20pt}\section{Cuts from Sequences of BCFW Deformations}\vspace{-14pt}

We would like to simplify the forward-limit term by recursing it in a way that will allow us to compute it analytically. Let us choose to label the \emph{first} loop momentum cut to be `$\ell_1$'. As the amplitude depends only on the \emph{line} $(A_1\,B_1)$, we may freely fix $A_1$ to be a special point on this line:
$A_1\!\mapsto\!\hat{\ell}_1\equivR\!(A_1\,B_1)\tncap(n\,\mi1\,n\,1)$---that is, $(A_1\,B_1)\simeq(\hat{\ell}_1\,B_1)$. Now consider a BCFW deformation which translates $B_1$ in the direction of $\hat{n}$---preserving the plane $(A_1\,B_1\,1)\simeq(A_1\,B_1\,\hat{n})$. 

With this deformation, physical poles will correspond to places where the deformed line $(\hat{\ell}_1\,B_1)$ intersects either an external line $(j\,\mi1\,j)$ or an internal line corresponding to another loop momentum, say $\ell_2\equivR\!(A_2\,B_2)$. The former results in the second term of (\ref{schematic_cut_formula}), and the latter corresponds to a diagram with two particles taken in the forward limit (two internal lines on-shell). To resolve the latter, we repeat this procedure by translating $A_2\!\mapsto\!\hat{\ell}_2$ and deforming $B_2$, and so-on, until all loop momenta have been exhausted---at which point the final $\ell_L$ can be deformed to expose one final `factorization' involving either an external or already-localized internal line $(\hat{\ell}_{a-1}\,\hat{\ell}_a)$.

This sequence of BCFW deformations and cut solutions is illustrated in \mbox{Figure \ref{cut_sequence_figure}}. Letting the $a$th loop momentum be encoded by the line $(A_a\,B_a)$ in twistor space, the cut points appearing in the $q$-loop bridge take the form summarized on the right hand side of \mbox{Figure \ref{cut_sequence_figure}}. 

A subtlety which will be clarified below is that the final line cut through this process need not involve any external momenta: the line `$(j\,\mi1\,j)$' may be chosen from any pair of adjacent elements of the list $(\hat{\ell}_{q{-}2},\hat{\ell}_{q{-}3},\ldots,\hat{\ell}_1,1,\ldots,n\,\mi1,\hat{n},\hat{\ell}_{1},\hat{\ell}_2,\ldots,\hat{\ell}_{q{-}2})$.

\begin{figure*}[t]
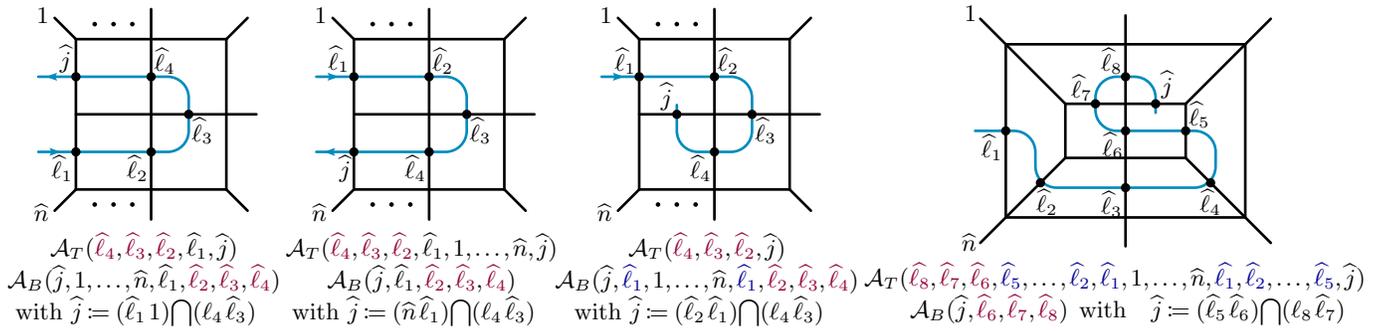
$$\fwbox{0pt}{\hspace{-0pt}\begin{array}{@{}c@{}}\\[-18.5pt]\hspace{-0pt}\fourLoopCutPathFigureA\\[03.75pt]\begin{array}{@{}c@{}}\\[-7.5pt]\mathcal{A}_T(\r{\hat{\ell}_4},\r{\hat{\ell}_3},\r{\hat{\ell}_2},\hat{\ell}_1,\hat{j})\\\mathcal{A}_B(\hat{j},1,\ldots,\hat{n},{\hat{\ell}_1},\r{\hat{\ell}_2},\r{\hat{\ell}_3},\r{\hat{\ell}_4})\\[0pt]
\fwboxR{15pt}{\text{with }}\hat{j}\equivR\!(\hat{\ell}_1\,1)\tncap(\ell_4\,\hat{\ell}_3)\\
\end{array}\end{array}\hspace{2pt}
\begin{array}{@{}c@{}}\\[-18.5pt]\hspace{-0pt}\fourLoopCutPathFigureB\\[03.75pt]\begin{array}{@{}c@{}}\\[-7.5pt]\mathcal{A}_T(\r{\hat{\ell}_4},\r{\hat{\ell}_3},\r{\hat{\ell}_2},\hat{\ell}_1,1,\ldots,\hat{n},\hat{j})\\\mathcal{A}_B(\hat{j},{\hat{\ell}_1},\r{\hat{\ell}_2},\r{\hat{\ell}_3},\r{\hat{\ell}_4})\\[0pt]
\fwboxR{15pt}{\text{with }}\hat{j}\equivR\!(\hat{n}\,\hat{\ell}_1)\tncap(\ell_4\,\hat{\ell}_3)\\
\end{array}\end{array}\hspace{-1pt}
\begin{array}{@{}c@{}}\\[-18.5pt]\hspace{-0pt}\fourLoopCutPathFigureC\\[03.75pt]\begin{array}{@{}c@{}}\\[-7.5pt]\mathcal{A}_T(\r{\hat{\ell}_4},\r{\hat{\ell}_3},\r{\hat{\ell}_2},\hat{j})\\\mathcal{A}_B(\hat{j},\b{\hat{\ell}_1},1,\ldots,\hat{n},\b{\hat{\ell}_1},\r{\hat{\ell}_2},\r{\hat{\ell}_3},\r{\hat{\ell}_4})\\[0pt]
\fwboxR{15pt}{\text{with }}\hat{j}\equivR\!(\hat{\ell}_2\,\hat{\ell}_1)\tncap(\ell_4\,\hat{\ell}_3)\\
\end{array}\end{array}\hspace{2pt}
\begin{array}{@{}c@{}}\\[-20.pt]\hspace{-0pt}\eightLoopCutPathFigure\\[03.75pt]\begin{array}{@{}c@{}}\\[-7.5pt]\mathcal{A}_T(\r{\hat{\ell_8}},\r{\hat{\ell}_7},\r{\hat{\ell}_6},\b{\hat{\ell}_5},\ldots,\b{\hat{\ell}_2},\b{\hat{\ell}_1},1,\ldots,\hat{n},\b{\hat{\ell_1}},\b{\hat{\ell}_2},\ldots,\b{\hat{\ell}_5},\hat{j})\\
\hspace{-95pt}\mathcal{A}_B(\hat{j},\r{\hat{\ell}_6},\r{\hat{\ell}_7},\r{\hat{\ell}_8})\fwboxL{0pt}{\;\;\text{with}\quad\hat{j}\equivR\!(\hat{\ell}_5\,\hat{\ell}_6)\tncap(\ell_8\,\hat{\ell}_7)}\\
\end{array}\end{array}
}\vspace{-15pt}$$\caption{Examples of novel cut sequences appearing in the recursion. The first two figures illustrate diagrams where one amplitude involves no external momenta while the third and fourth figures illustrate terms where one side of the bridge is entirely contained within the other.\\[-16pt]\label{example_cut_sequences_figures}}\end{figure*}

\vspace{-0pt}\subsection{Explicit form of the `Bridged' Amplitudes Resulting from Recursion}\vspace{-14pt}
%
%
When $q$ internal loop momenta are cut via the sequence of BCFW deformations described above, the resulting residue of the loop integrand takes the form
\vspace{-10pt}\eq{\begin{split}
~\\[-6pt]\fwbox{0pt}{\hspace{-30pt}\bridgeDefnFigure}\hspace{20pt}\bigger{}\fwboxL{150pt}{\hspace{00pt}\begin{array}{@{}l@{}l@{}}~\\[15pt]
\equivR\!\!\!\displaystyle\int\!\!\!\prod_{a{=}2}^qd^4\!\r{\eta_{\hat{\ell}_a}}\Big\{&\mathcal{A}_T^{L_T}(\r{\hat{\ell}_q}\,\r{\hat{\ell}_{q{-}1}}\r{\hat{\ell}_{q{-}2}}\cdots\hat{j})\\[-5pt]
&\hspace{-0pt}\mathcal{B}_q[1,(j\,\mi1\,j),(n\,\mi1\,n)]\\
&\mathcal{A}_B^{L_B}(\hat{j}\cdots\r{\hat{\ell}_{q{-}2}}\,\r{\hat{\ell}_{q{-}1}}\r{\hat{\ell}_{q}})\,\,\Big\}\end{array}}\hspace{-20pt}\\[-6pt]\end{split}\vspace{-5pt}\label{detailed_expression_for_bridged_amps}}
where $L_T{+}L_B{=}L{-}q$ and the bridge factors `$\mathcal{B}_q$' are derived by performing the ${\rm GL}_2$-integrals required for the forward-limit (see \cite{ArkaniHamed:2010kv}) within products of $R$-invariants.
The 0-loop bridge $\mathcal{B}_0$ was given above in (\ref{zero_bridge}); for $q\!=\!1$ it is
\vspace{5pt}\eq{\begin{split}
\\[-10pt]\hspace{-45pt}&\hspace{-20pt}\mathcal{B}_1[1,\b{(j\,\mi1\,j)},\g{(n\,\mi1\,n)}]\equivR\\
&\hspace{-20pt}\frac{\ab{\ell_1(\b{j\,\mi1\,j}\,1)\tncap(\g{n\,\mi1\,n}\,1)}^2}{\ab{\ell_1\,\b{j\,\mi1\,j}}\ab{\ell_1\,\b{j}\,1}\ab{\ell_1\,1\,\b{j\,\mi1}}\ab{\ell_1\,\g{n\,\mi1\,n}}\ab{\ell_1\,\g{n}\,1}\ab{\ell_1\,1\,\g{n\,\mi1}}},\hspace{-20pt}\\[-4pt]\end{split}\vspace{-10pt}}
which appears as the `kermit' of refs.~\cite{ArkaniHamed:2010gg,Bourjaily:2013mma}; and for $q\geq2$ we obtain the new result
\begin{widetext}
\vspace{-5pt}\eq{\hspace{-35pt}\underset{\fwboxL{0pt}{\hspace{-10pt}(q\geq2)}}{\mathcal{B}_q}[1,(j\,\mi1\,j),(n\,\mi1\,n)]\equivR\frac{\ab{(n\,\mi1\,n\,1)\,\r{\hat{\ell}_2}}\Big(\!\ab{\hat{\ell}_0\hat{\ell}_1\r{\hat{\ell}_2}\r{\hat{\ell}_3}}\ab{\hat{\ell}_1\r{\hat{\ell}_2}\r{\hat{\ell}_3}\r{\hat{\ell}_4}}\cdots\ab{\r{\hat{\ell}_{q{-}3}}\r{\hat{\ell}_{q{-}2}}\r{\hat{\ell}_{q{-}1}}\r{\hat{\ell}_{q}}}\ab{\r{\hat{\ell}_{q{-}2}}\r{\hat{\ell}_{q{-}1}}\r{\hat{\ell}_{q}}\hat{j}}\!\Big)\ab{\r{\hat{\ell}_{q{-}1}}\r{\hat{\ell}_{q}}(j\,\mi1\,j)}}{\ab{\ell_1\,n\,\mi1\,n}\ab{\ell_1\,n\,1}\ab{\ell_1\,1\,n\,\mi1}\big(\ab{\ell_1\,\ell_2}\cdots\ab{\ell_{q{-}1}\ell_q}\big)\ab{\ell_q\,j\,\mi1\,j}\ab{\ell_q\,j\,\hat{\ell}_{q{-}1}}\ab{\ell_q\,\hat{\ell}_{q{-}1}\,j\,\mi1}}.\hspace{-20pt}\label{L_loop_bowtie_formula}\vspace{-5pt}}
\end{widetext}
Notice that the lower-loop amplitudes being bridged have an excess $R$-charge of $4(1{-}q)$, which is compensated by the $(q-1)$ $d^4\eta_{\hat{\ell}_a}$ Grassmann integrations in \eqref{detailed_expression_for_bridged_amps}. It is worth emphasizing that the expression \eqref{detailed_expression_for_bridged_amps} is exactly dual-conformal (although not manifest in the expression given in (\ref{L_loop_bowtie_formula})): the factors of $\mathcal{B}_q$ carry net weight ${+}4$ in $\r{\hat{\ell}_2},\ldots,\r{\hat{\ell}_q}$ (which is compensated by the $\r{\eta_{\hat{\ell}_a}}$-integrations) and weight ${-}4$ in $\ell_1,\ldots,\ell_q$. 

Finally, all results must be symmetrized over the $L$ lines $\{(A_1B_1),\ldots (A_L B_L)\}$.

\vspace{-12pt}\section{Novelty Relative to Ordinary `Unitarity Cuts'}\vspace{-14pt}
%
Although a propagator involving each loop momentum is cut sequentially in the recursion, poles resulting from the final BCFW deformation (of $\ell_q$ within the plane $(A_{q{-}1}\,B_{q{-}1}\,\hat{\ell}_{q{-}2})$) need not involve an `external' propagator: in addition to the `external' poles of the form $\ab{A_q\,B_q\,j\,\mi1\,j}$, there can also be those involving the lines  $(\hat{\ell}_1\,1)$ or $(\hat{n}\,\hat{\ell}_1)$, or even the purely `internal' poles of the form $\ab{A_q\,B_q\,\hat{\ell}_{a{-}1}\,\hat{\ell}_{a}}$ for any $a\leq (q{-}2)$.

Examples of the first type of novelty are illustrated in the first two figures in \mbox{Figure \ref{example_cut_sequences_figures}}, respectively. These yield a product of two amplitudes in which one of the factor only contains internal lines.  This is possible since, in contrast with the Cutkosky rules \cite{Cutkosky:1960sp}, there is no requirement that all energies in our complexified cuts be positive. These first appear at three loops as all amplitudes being bridged must involve at least four legs.

The second type of novelty is illustrated in the last two examples of \mbox{Figure \ref{example_cut_sequences_figures}}, where an internal propagator is cut at the final stage of the recursion and one finds a completely internal island on one side of the bridge. This `cut' is unquestionably unusual and in fact involves one amplitude with at least one pair of (necessarily non-adjacent) legs taken in the forward-limit. This may seem problematic, since the complexity involved in evaluating forward limits required by the recursion formula of \cite{ArkaniHamed:2010kv} was a primary technical obstruction motivating our present work. Nevertheless, we find that non-adjacent forward limits are rather benign and that any term that would require delicacy vanishes directly upon $d^4\r{\eta_{\hat{\ell}_a}}$-integration. Thus, all terms with the requisite $\r{\eta_{\hat{\ell}_a}}$-support may be evaluated na\"ively in the forward-limit. 

It is interesting to count the number of terms $N({n,k,L})$ generated by applying the recursion (\ref{schematic_cut_formula}) for the $n$-point N${}^k$MHV amplitude at $L$-loops.
Defining a generating function $\mathcal{G}(\nu,\kappa,\lambda)\equivL\sum_{n,k,L} N({n,k,L}) \nu^n \kappa^k \lambda^L$ with $n\geq 4$, the recursion implies that 
\eq{\mathcal{G}=\nu^4{+}\nu(1{+}\kappa)\mathcal{G}{+}\left.\left(\frac{\kappa^2\mathcal{G}^2}{\nu^2\,\kappa{-}\lambda}\right)\right|_{\text{reg.}}\,,}
where the second term accounts for tree-level bridges involving a three-point vertex of either parity, and the quadratic term with $\frac{\kappa^2}{\nu^2\kappa-\lambda} =\frac{\kappa}{\nu^2}{+}\frac{\lambda}{\nu^4}{+}\frac{\lambda^2}{\nu^6\kappa}{+}\ldots$ accounts for all the other bridges.  The `reg.' operation removes any terms with negative or too-large powers of $\kappa$
so as to maintain the N${}^k$MHV degree within the range $k\!\in\![0,n{-}4]$---terms outside this range vanish trivially upon Grassmann integration in (\ref{detailed_expression_for_bridged_amps}).
The solution for $\lambda{=}0$ gives the Catalan numbers familiar from tree-level recursion (see e.g.~\cite{ArkaniHamed:2012nw}), while the one-loop counting agrees with \cite{ArkaniHamed:2012nw}. We do not know a closed form for $N({n,k,L})$ for $L\geq2$, but these numbers agree with the number of terms produced by the function \texttt{preAmp[]} in the attached package.

We found that a significant (and recursion-scheme-dependent) fraction of these terms vanish upon Grassmann integration. At three loops, for example, we have found expressions for the 4-particle amplitude involving anywhere between 88--98 non-vanishing terms after Grassmann integration (significantly fewer than the 146 terms prior to Grassmann integration). In the included \textsc{Mathematica} package, we default to a recursion scheme wherein non-cut-loop variables are preferentially chosen for subsequent recursion; but we have also included a function \texttt{superAmpRandom[]} which randomly selects the legs to be BCFW-deformed at each step of the recursion. The agreement between the different resulting expressions is a powerful consistency check on our implementation of the all-loop formula (\ref{schematic_cut_formula}).

\vspace{-16pt}\section{Conclusions and Discussion}\vspace{-14pt}
%
In this note, we have used a sequence of BCFW deformations to arrive at novel, recursive representation of any $L$-loop, $n$-point N${}^k$MHV amplitude in planar, maximally supersymmetric ($\mathcal{N}\!=\!4$) Yang-Mills theory in terms of bridges between lower-loop amplitudes in the theory. By avoiding any reference to problematic forward-limits, the formula given in (\ref{schematic_cut_formula}) can be directly used to evaluate amplitude integrands well beyond the reach of existing, general algorithms. We have implemented these tools in a relatively simple \textsc{Mathematica} package, included among the ancillary files of this work. (These tools have been built using those included in the works of \cite{Bourjaily:2010wh,Bourjaily:2013mma,Bourjaily:2015jna}.) We have explicitly checked that the results for our integrands match the local expressions for four points through five loops (see e.g.~\cite{Bern:1997nh,Bern:2004ne,Bern:2005iz,Bern:2006ew,Bern:2007ct}). 

It has been conjectured that all multiplicity, lower-loop amplitudes can be determined via an $n$-point light-like limit of the 4-point correlation function at sufficiently high loops \cite{Alday:2010zy,Eden:2010zz,Eden:2010ce,Eden:2011yp,Adamo:2011dq,Eden:2011ku,Heslop:2018zut}; in the case of 5 particles, this encoding is explicit,
giving the complete $L$-loop integrand in terms of the $(L{+}2)$-loop 4-point correlator (this includes an extra loop needed to tease out the parity-odd part of the integrand (see e.g.~\cite{Eden:2011ku,Heslop:2018zut})). We have checked this result explicitly through 5 loops, which can be viewed as both a check on the correctness of the $n$-point projection from the $4$-point correlator and on the correlator of stress tensors itself, as taken from \cite{Bourjaily:2011hi,Bourjaily:2015bpz,Bourjaily:2016evz}. 

Although transparent and relatively efficient, it is worth noting that---unlike BCFW applied to amplitudes at tree-level---the representations that result from (\ref{schematic_cut_formula}) are \emph{far} from compact, dramatically exceeding the number of terms required in `local' integrand formulae such as those of \cite{Bourjaily:2013mma,Bourjaily:2015jna}, for example. The four-particle integrand at 5 loop-order (before loop label symmetrization) is given by 23,072 terms by our package (using default options for how bridged amplitudes are recursed); this is far greater than representations of this amplitude in terms of local integrands. For example, even including the dihedral images of the 34 archetype terms of the representation of \cite{Bern:2007ct} the local loop integrand representation involves a mere 193 terms.

This `inefficiency' is easy to understand: each term appearing in (\ref{schematic_cut_formula}) exposes a very specific sequence of physical cuts whereas a single local integrand can capture the contributions from many individual cuts simultaneously. This inefficiency may reflect the eventual tension between the singularity structure of particular amplitudes and the locality of individual Feynman integrals: our representation should represent a triangulation of the amplituhedron, related to a particular `$d\!\log$'-form \cite{Arkani-Hamed:2014via} in the Grassmannian; as such, it might do a better job of exposing the analytic structure of amplitudes (or even perhaps be easier to integrate---once IR regularization and loop integration are understood for these expressions \cite{Lipstein:2013xra}) than expressions involving Feynman-diagram-like master integrands. 

It would be interesting to know if a formula such as (\ref{schematic_cut_formula}) could be applied to less than maximally supersymmetric theories, or to maximally supersymmetric theories beyond the planar limit. 

\nopagebreak

\vspace{-12pt}\section{Acknowledgments}\vspace{-15pt}
The authors gratefully acknowledge fruitful conversations with Jaroslav Trnka. 
This project has been supported by an ERC Starting Grant \mbox{(No.\ 757978)}, a grant from the Villum Fonden \mbox{(No.\ 15369)}, and the US Department of Energy under contract DE-SC00019066 (JLB); and by the Simons Collaboration on the Nonperturbative Bootstrap, the Simons Fellowships in Theoretical Physics, and the Canada Research Chair program (SCH).

%
\providecommand{\href}[2]{#2}\begingroup\raggedright\endgroup

\end{document}